# Interferometric apodization by homothety – II. Experimental validation

J. Chafi ,[1,2]★ Y. El Azhari,[1,2,3]★ O. Azagrouze,[1,2,3] A. Jabiri,[1,2] A. Boskri ,[1,2] Z. Benkhaldoun [1,2] and A. Habib[1,2,3]

[1]LPHEA, Faculté des Sciences Semlalia, Université Cadi Ayyad, Av. Prince My Abdellah, BP 2390 Marrakech, Morocco
[2]Oukaimeden Observatory, Cadi Ayyad University, 40273 Marrakech, Morocco
[3]Centre Régional des Métiers de l'Education et de la Formation de Marrakech, 40000 Marrakech, Morocco

## ABSTRACT

This work presents the results of experimental laboratory tests on the apodization of circular and rectangular apertures using the Interferometric Apodization by Homothety (IAH) technique. The IAH approach involves splitting the amplitude of the instrumental PSF into two equal parts. One of the two produced PSFs undergoes homothety to change its transverse dimensions while its amplitude is properly controlled. The two PSFs are then combined to produce an apodized image. The diffraction wings of the resulting PSF are subsequently reduced by some variable reduction factor, depending on an amplitude parameter $\gamma$ and a spread parameter $\eta$. This apodization approach was implemented in the laboratory using an interferometric set-up based on the Mach–Zehnder Interferometer (MZI). The experimental results exhibit a strong agreement between theory and experiment. For instance, the average experimental contrast obtained at a low angular separation of $2.4\lambda/D$ does not exceed $5 \times 10^{-4}$. This work also allowed us to study the influence on the apodizer's performance of some parameters, such as the wavelength and the density of the neutral filters.

## 1 INTRODUCTION

In high-dynamic range imaging, apodization is a commonly used technique for enhancing the performance of coronagraphs (Soummer 2005; Soummer et al. 2009). In the field of direct observation of exoplanets, the utility of apodized pupils has been studied by Nisenson & Papaliolios (2001). Aime, Soummer & Ferrari (2001) suggested using interferometry for pupil apodization, and later, Aime, Soummer & Ferrari (2002); Soummer, Aime & Falloon (2003) showed that the rejection produced by a Lyot coronagraph could be improved by apodizing the entrance pupil with a prolate spheroidal function.

Subsequently, Soummer (2005) demonstrated that for a random pupil, solutions based on prolate functions could also checked the performance of the Lyot coronagraph. Aime (2005) proposed apodizing a circular aperture using an MZI and employing thin lenses with opposite vergence to achieve apodization. This apodization technique was experimentally verified by El Azhari et al. (2005) and Azagrouze, El Azhari & Habib (2008) using a Michelson interferometer (MI), and also by Carlotti et al. (2008) using a MZI. Furthermore, N'diaye et al. (2012) explored the application of an apodization concept, the 'Four Quadrant Zeroth Order Grating (4QZOG)', developed by Mawet et al. (2005), in coronagraphy. They demonstrated that using coloured apodization applied before a 'Dual Zone Phase Mask

(DZPM)' coronagraph can enhance the contrast performance over a broader spectral band, thereby optimizing performance in the presence of various types of noise.

The grey-scale apodizers were converted into binary masks for manufacturing reasons, leading to the emergence of Apodized Pupil Lyot Coronagraphs (APLC; N'Diaye et al. 2016; Zimmerman et al. 2016), used in instruments such as VLT/SPHERE (Guerri et al. 2011) and Gemini/GPI (Sivaramakrishnan et al. 2010). These coronagraphs are also planned for future space telescopes such as the *Roman Space Telescope* (*RST*; Zimmerman et al. 2016), LUVOIR-A, and LUVOIR-B (Stahl, Shaklan & Stahl 2015; Leboulleux et al. 2017; Stahl 2017; Leboulleux et al. 2018a; Leboulleux et al. 2018b; Laginja et al. 2019; The LUVOIR Team 2019; Laginja et al. 2020; Stahl, Nemati & Stahl 2020; Laginja et al. 2021; Juanola-Parramon et al. 2022). Second-generation ground-based imagers, such as the Planetary Camera and Spectrograph (PCS) for the ELT (Kasper, Verinaud & Mawet 2013), the Planetary Systems Imager (PSI) for the Thirty Meter Telescope (TMT; Fitzgerald et al. 2019) and GMagAO-X for the Giant Magellan Telescope (GMT; Males et al. 2018) will target fainter and closer planets to their host stars, ranging from young Jupiter's to terrestrial planets. The Astro2020 Decadal Survey by NASEM (National Academies of Sciences, Engineering, and Medicine 2021) also supports the development of large space telescopes to observe Earth-like exoplanets, such as LUVOIR-B (The LUVOIR Team 2019) and Habitable Exoplanet (HabEx; Gaudi et al. 2020).

Carlotti (2013) and Carlotti, Pueyo & Mawet (2014) focused their research on the apodization of vortex phase mask coronagraphs for arbitrary apertures, especially axially obstructed circular apertures (including spiders). They stated that apodizers could be optimized for the application of phase masks on these apertures and compared these results to those of E-VLT type apertures.

Habib et al. (2010) and Azagrouze et al. (2010) introduced a new approach to interferometric apodization, called IAH technique, which utilizes homothety. This method yields results similar to those obtained with a circular aperture using step-like transmission. We refer to this transmission as being obtained using two nested circular symmetric step functions with respective aperture radii $R_1$ and $R_2 < R_1$. The IAH technique efficiently achieves very high dynamic imaging by minimizing diffraction noise in the final images. Specifically, 93.6 per cent of the stellar light is concentrated within the central diffraction spot. This could make it a valuable complement to commonly used coronagraphs in major observatories, such as APLC, Vortex, and others, enhancing their performance.

The IAH technique is based on duplicating the original PSF to generate two identical PSFs via amplitude division. These PSFs are subsequently coherently superimposed. The core strategy is to stretch one PSF and modulate its amplitude, ensuring that the negative lobes of one align with the positive lobes of the other, effectively reducing the diffraction wings. This method facilitates versatile modulation of the apodization to suit diverse configurations and needs.

One of the many advantages of the IAH technique is its ease of implementation, as it does not require complex equipment and can be easily integrated into existing coronagraphs. Furthermore, this approach can be adapted to cascade multiple stages of IAH, thereby enhancing the rejection of secondary PSF lobes and improving coronagraph performance. However, using the IAH technique in cascading stages presents a distinct trade-off. On one hand, it enhances the efficiency of the apodization, offering improved performance, such as the reduction of diffracted light. On the other hand, it increases the difficulty of adjustments and leads to signal losses, which can potentially degrade the signal-to-noise ratio. It also uses simple achromatic optical elements such as apertures and neutral density filters, streamlining the apodizer assembly and reducing chromatic aberrations. The IAH technique holds great potential as a powerful tool for obtaining High Dynamic Range images and enhancing the performance of existing coronagraphs.

In our previous article (Chafi et al. 2023), we presented simulation results for various aperture shapes, such as circular, rectangular, hexagonal, and apertures similar to those used in the TMT telescope. These results demonstrated promising performance in enhancing coronagraph capabilities by increasing contrast at low-angular separations, making the IAH technique even more attractive for use in major observatories. In the current paper, we aim to further this work by presenting the experimental results of IAH technique on circular and rectangular apertures.

While simulation results are highly valuable for understanding the theoretical aspects of our IAH method, experimental validation is crucial to prove its practical feasibility. Specifically, laboratory testing of the IAH technique allows us to verify simulation results and, most importantly, assess the sensitivity of various optical element imperfections and sources of noise. Hence, experimental validation is essential to confirm the effectiveness of our technique and establish its credibility for practical applications. The experimental results will also enable more precise calibration of simulation models to improve performance prediction under real-world conditions.

The experimental implementation of our study involves many key challenges. First, it requires delicate and precise manipulations, including meticulous adjustment of the MZI, as well as rigorous control of essential components for the IAH technique, such as circular and rectangular apertures, and neutral density filters in the MZI arms. Accomplishing these manipulations necessitates the use of specialized equipment, strict adherence to a protocol, and meticulous attention to detail to minimize experimental errors. It is important to emphasize that the inherent difficulty in this experimental implementation is unavoidable due to our study's ambitious and innovative nature. In fact, the IAH technique itself is based on a very simple (non-complex) principle. However, in the experimental implementation, we propose, we have used a MZI, the alignment of which, especially in polychromatic light, is not easy.

We arranged this paper as follows: in Section 2, we provide a detailed description of the experimental set-up used to implement the IAH technique. Next, in Section 3, we present the experimental results. We started by exploring the case of a rectangular aperture before moving on to a circular one. In the same context, we took the opportunity to examine the chromatic behaviour of the IAH technique using various wavelengths. Finally, Section 4 summarizes our findings and outlines our prospects for future research.

## 2 EXPERIMENTAL SET-UP

The choice of the experimental set-up is of paramount importance in the realization of IAH. The objective of this section is to provide a concise yet sufficiently detailed description of the optical set-up used, as well as the protocol applied for image acquisition.

### 2.1 Optical assembly

The experimental set-up used for the IAH technique is depicted in Fig. 1. It incorporates a monochromatic He–Ne laser with a wavelength of 543.5 nm, featuring a non-adjustable cavity. To assess the chromatic behaviour of our system, we substituted this laser with others, whose wavelengths will be detailed in Section 3.4, each operating at a unique wavelength. This enabled us to evaluate the efficacy of our device across various wavelengths, a 5 × microscope objective (OC) with a focal length of 25.4 mm, and a 100 mm lens ($L_0$) to produce cylindrical beams. A folding mirror ($M_0$) is used to optimize the workspace on the Smart Table M-OTS-UT2 optical table. Our experiment relies on the use of the MZI, assembled with two semireflective beam splitters, $BS_1$ and $BS_2$, and two plane mirrors, $M_1$ and $M_2$ (Fig. 2). The MZI has two outputs: one with in-phase beams (additive output), allowing the observation of the apodization effect, and the other with out-of-phase beams (subtractive output), resulting in an anti-apodization effect. In this case, the energy is redirected towards the secondary lobes. This duality in the MZI's outputs serves as a crucial component of our validation process. It not only verifies the accuracy of the apodization but also facilitates the measurement of the total light source intensity. This measurement is essential for precisely determining the intensity of an exoplanet, optimizing our results, and enhancing the characterization of exoplanets. The spatial arrangement of the different elements in the MZI facilitates the introduction of the necessary elements for the IAH technique. The set-up is designed to operate with single circular apertures or rectangular apertures, depending on the desired apodization type.

To achieve circular symmetry apodization, we used two circular apertures, $\mathcal{P}_1$ and $\mathcal{P}_2$, with respective diameters $\phi_1$ and $\phi_2$, allowing us to control the beam diameters with a homothety ratio $\eta = \frac{\phi_2}{\phi_1}$. The beam intensity control was accomplished using two neutral density filters, $ND_1$ and $ND_2$. Both apertures and filters were carefully positioned in the two arms of an MZI, as illustrated in Fig. 1. A

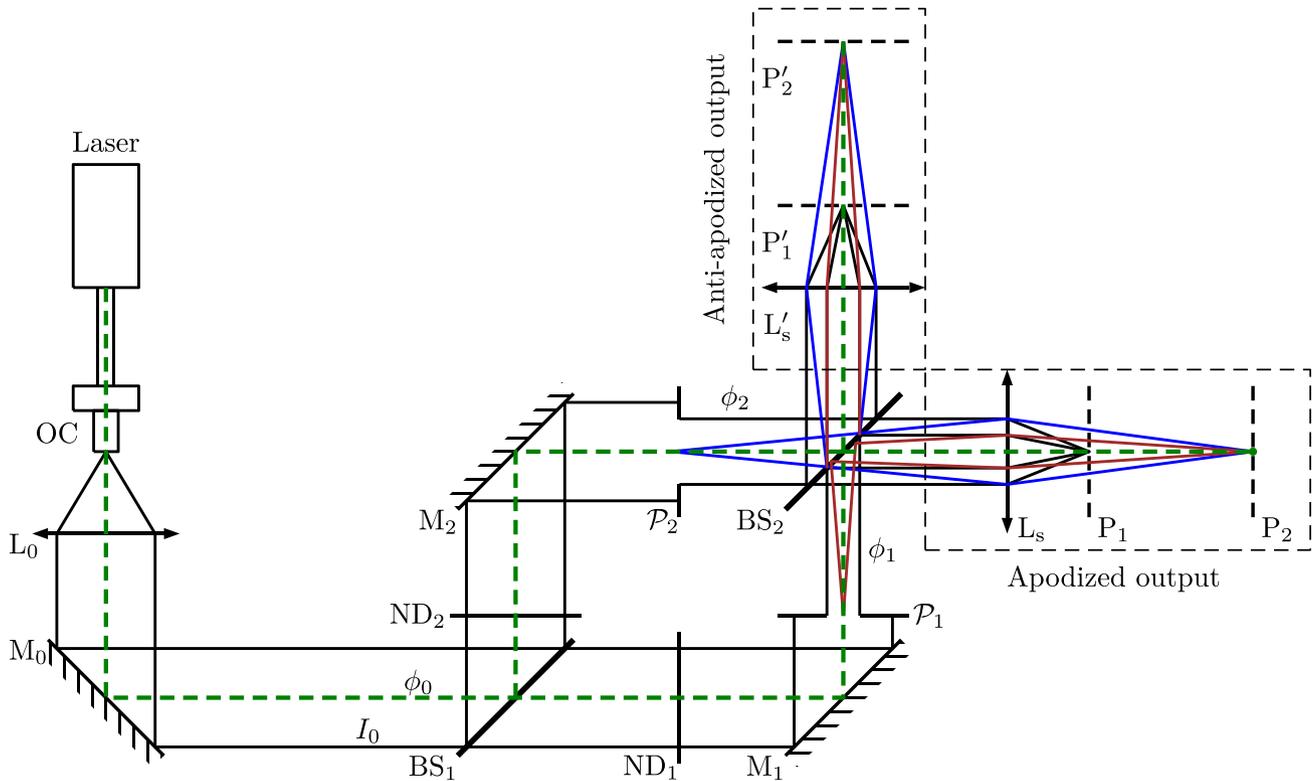

**Figure 1.** The optical set-up, based on the MZI, was designed to operate with either single circular apertures or adjustable rectangular apertures, depending on the desired apodization type. The set-up comprises two semireflective beam splitters, $BS_1$ and $BS_2$, along with two plane mirrors, $M_1$ and $M_2$. Two single circular apertures, $\mathcal{P}_1$ and $\mathcal{P}_2$, with respective diameters $\phi_1$ and $\phi_2$ (or two rectangular apertures) were used. Two neutral density filters, $ND_1$ and $ND_2$, enabled beam intensity control. The diameter of the incident beam with intensity $I_0$ is denoted as $\phi_0$. The final focal plane to observe the apodization effect on the PSF is $P_1$, while the image plane to observe the apodized pupil is $P_2$.

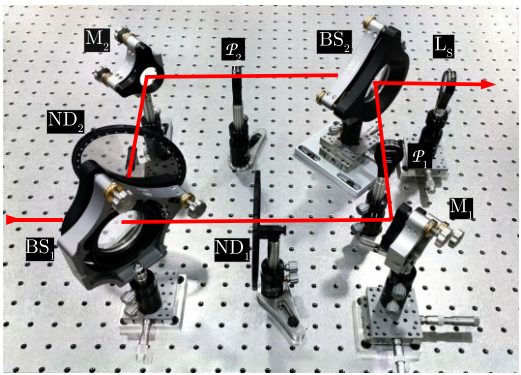

**Figure 2.** Image illustrating the interferometric part of the experiment, implementing interferometric apodization through homothety. The light beams go into the MZI from the left and exit from the right. Two circular or rectangular apertures, $\mathcal{P}_1$ and $\mathcal{P}_2$, are accurately positioned between the second beam splitter and the two side mirrors to control the beam diameters. Two neutral density filters, $ND_1$ and $ND_2$, are placed between the side mirrors and the first beam splitter to control the intensities of the two beams. The notations used for the mentioned components are the same as those used in Fig. 1.

200 mm lens ($L_s$) formed the image of the apodized aperture on the final image plane $P_2$ of the instrument, allowing the observation of the apodization effect in the focal plane $P_1$ of $L_s$. Both planes were observed successively by placing the CCD camera in the appropriate plane.

## 2.2 Image acquisition and processing

To capture the obtained images, we employed a high-resolution imaging camera equipped with a KAF-8300 CCD sensor. This sensor boasts a surface area of $17.96 \times 13.52\,\mathrm{mm^2}$ and features a $3326 \times 2504$ pixels array, resulting in over 8.3 megapixels with each pixel measuring 5.4 μm on each side. An important attribute for acquiring apodized PSF images is the sensor's dynamic range or bit depth. The CCD camera used has a 16-bit dynamic range, enabling image encoding with more than 65 000 levels of grey-scale.

To mitigate the dark current and read out noise, the camera is equipped with an internal thermoelectric cooling system, capable of lowering the sensor's temperature to as low as $-40\,^\circ$C relative to the ambient temperature. The exposure time can be set between 0.1 and 60 min. However, we consistently opted for a minimal exposure time and adjusted the beam intensity to avoid pixel saturation. Subsequently, we performed extensive image processing to correct for dark current effects using the AstroImageJ interface (Collins et al. 2017, AIJ).

## 3 EXPERIMENTAL RESULTS

In this section, we present the experimental results of the IAH technique obtained in the laboratory using the MZI-based set-up depicted in Fig. 1 and described in Section 2. We will sequentially present the results obtained with a rectangular aperture and then those obtained with a circular aperture. Additionally, we will explore the influence of various parameters on the performance of the IAH technique. Specifically, we will investigate the impact of the wavelength of the light used. This will provide insights into the chromatic response or chromatism of the IAH technique.

### 3.1 Rectangular aperture

We started by experimentally studying the case of a rectangular aperture. The advantage of this configuration lies primarily in the availability of high-quality rectangular apertures from optical suppliers, with dimensions that can be easily and accurately controlled in a continuous manner. This simplifies the set-up process significantly, streamlining the adjustments needed for experimentation.

The use of this geometric aperture shape provides the advantage of reducing secondary diffraction lobes along both (*x* and *y*) axes and, more importantly, along the diagonals, as demonstrated by Zanoni & Hill (1965). Nisenson & Papaliolios (2001) reproduced an apodized rectangular aperture using transmission functions that create extended regions around the PSF diagonals where the intensity decreases rapidly. More recently, Itoh & Matsuo (2022) have explored the use of a rectangular aperture to achieve deep nulling.

Before presenting the experimental results of rectangular apertures, let us recapitulate some steps involved in optimizing the design of the experimental set-up. These steps have been detailed elsewhere in a previous article (Chafi et al. 2023).

The IAH technique involves superimposing in the instrument's focal plane two PSFs that are coherent with each other, where one is judiciously stretched (with a stretching factor $\eta$) and its amplitude is affected by a multiplicative factor $\gamma$. The resulting amplitude at the MZI output in the in-phase channel is then given by:

$$A(u, v, \gamma, \eta) = \frac{\sin(\pi u)}{\pi u} \frac{\sin(\pi v)}{\pi v} + \gamma \frac{\sin(\pi u \eta)}{\pi u \eta} \frac{\sin(\pi v \eta)}{\pi v \eta}. \quad (1)$$

The out-of-phase channel produces an anti-apodized amplitude. To obtain this amplitude, we simply replace the parameter $\gamma$ with $-\gamma$ in equation (1).

These amplitudes are coherent superpositions, respectively additive and subtractive, of the Fourier transforms of the transmittances $t_1(x, y)$ and $t_2(x, y)$ of two rectangular apertures with respective dimensions $(\ell_x, \ell_y)$ and $(\eta \ell_x, \eta \ell_y)$, each placed in one of the two arms of the MZI:

$$t_1(x, y) = \Pi\left(\frac{x}{\ell_x}\right) \Pi\left(\frac{y}{\ell_y}\right) \quad \text{and}$$

$$t_2(x, y) = \gamma \Pi\left(\frac{x}{\eta \ell_x}\right) \Pi\left(\frac{y}{\eta \ell_y}\right). \quad (2)$$

Where $\Pi(\frac{x}{\ell_x})$ is the window function, equal to 1 for $-\frac{\ell_x}{2} < x < \frac{\ell_x}{2}$, and 0 otherwise.

The effectiveness of the apodization greatly depends on the values assigned to the parameters $\gamma$ and $\eta$. In a recent work (Chafi et al. 2023), we determined the optimal values for these two parameters to concentrate the maximum energy in the central lobe of the PSF. The optimal values are $\gamma = 1.928$ and $\eta = 1.657$. This corresponds to concentrating 91.8 per cent of the total energy in the central diffraction lobe, compared to 81.5 per cent without apodization. These values constrain the useful surface areas $S_1$ and $S_2$ of the two rectangular apertures $\mathcal{P}_1$ and $\mathcal{P}_2$, as well as the densities $D_1$ and $D_2$ of the two neutral density filters $ND_1$ and $ND_2$, through the following two relations:

$$S_2 = \eta^2 S_1 \quad \text{and} \quad \Delta D = D_1 - D_2 = 2 \log_{10}\left(\frac{\gamma}{\eta}\right). \quad (3)$$

Table 1 summarizes the optimal values of parameters $\gamma$ and $\eta$, along with the corresponding density difference value $\Delta D$. It also provides a set of possible values for the dimensions $\ell_1$ and $\ell_2$ of the square apertures to be used.

**Table 1.** Optimal values of the parameters $\gamma$ and $\eta$ for the rectangular geometry, along with the corresponding density difference $\Delta D$. $\ell_1$ and $\ell_2$ represent the dimensions of square apertures satisfying the optimal apodization conditions.

| $\gamma$ | $\eta$ | $\Delta D$ | $\ell_1$(mm) | $\ell_2$(mm) |
|---|---|---|---|---|
| 1.928 | 1.657 | 0.132 | 1 | 1.657 |

To approximate the apodization parameters $\gamma$ and $\eta$ as closely as possible to their optimal values $\gamma = 1.928$ and $\eta = 1.657$, we initially used commercially available neutral density filters with respective densities of $D_1 = 0.2$ and $D_2 = 0.1$. Additionally, we employed adjustable rectangular apertures and set the dimension of aperture $\mathcal{P}_1$ to $\ell_1 = \ell_1' (1 \pm 0.05)$ mm (square geometry). Subsequently, we conducted experimental adjustments to the dimension of aperture $\mathcal{P}_2$ to maximize the apodization in the $\mathcal{P}_1$ plane along both transverse directions. This was achieved with $\ell_2 = \ell_2' (1.63 \pm 0.15)$ mm, corresponding to experimental values of $\gamma_{exp} = 1.83$ (5 per cent deviation) and $\eta_{exp} = 1.63$ (2 per cent deviation). The average overall discrepancy between our theoretical predictions and experimental measurements is 3.5 per cent, thereby illustrating the closeness of our experimental results to the theoretical ones.

Fig. 3 provides an overview of the pupil plane (column a) and the focal plane (column b) for the cases of an unapodized pupil (top), an apodized pupil (middle), and an anti-apodized pupil (bottom). These images give a qualitative appreciation of the effect of IAH on the output in-phase (or apodized output) and output in opposition to the phase (or anti-apodized output) of the MZI, despite the slight differences between the experimental values of the $\gamma$ and $\eta$ parameters of the IAH set-up and the theoretical optimal values.

In order to quantify the apodization effect, we performed cross-sections along specific directions of the pupil plane image (Fig. 3a) and the focal plane image (Fig. 3b).

Fig. 4, on the other hand, provides intensity cross-sections in the pupil plane displayed on a linear scale in blue compared with the theoretical curve in red dashed lines. It is noteworthy that there is good agreement between the theoretical prediction and the experimental results. Indeed, the results predicted by Azagrouze (2012) are confirmed, stating that the IAH is equivalent to a stepwise transmission.

Fig. 5 shows the intensity cross-sections of the PSFs obtained in the focal plane along the *x*-axis. The experimental curves are displayed in blue, while the theoretical ones are shown in red. Overall, for both the unapodized case (a) and the apodized (b) and anti-apodized (c) cases, there is a good agreement between the experimental results and the theoretical predictions. The apodized PSF exhibits a slight broadening of the central lobe compared to the unapodized PSF. Conversely, a slight narrowing of the central lobe is observed for the anti-apodized PSF. The attenuation of secondary diffraction lobes, theoretically predicted (Chafi et al. 2023) is confirmed experimentally. This translates to an average contrast of $10^{-3}$ at a distance between $3\lambda/D$ and $4\lambda/D$. These results are comparable to those obtained by other apodization techniques, such as the one studied by El Azhari et al. (2005) and Azagrouze, El Azhari & Habib (2008), who experimentally validated the interferometric apodization of a one-dimensional rectangular aperture using a Michelson interferometer. Similarly, Carlotti et al. (2008) performed apodization of a rectangular aperture using the MZI.

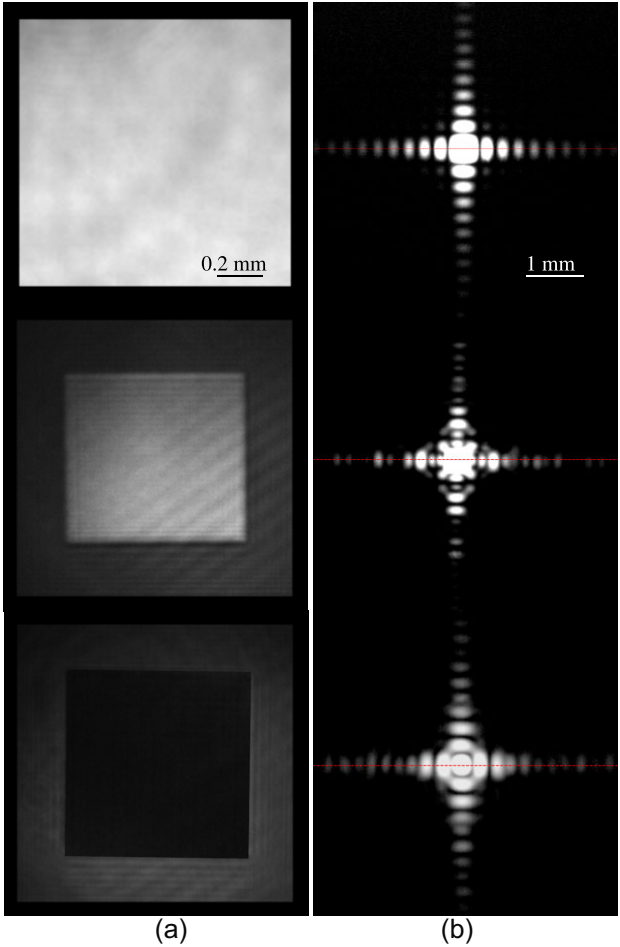

0.2 mm

1 mm

(a)                    (b)

**Figure 3.** Rectangular apodization: column (a) represents the pupil plane, while the column (b) displays the focal plane. At the top, we observe the case of an unapodized pupil; in the middle, an apodized pupil; and at the bottom, an anti-apodized pupil. The focal plane intensities are presented on a logarithmic scale, and the corresponding theoretical expressions are given by equation (1) for the following three cases: the unapodized case ($\gamma = 0$), the apodized case ($\gamma$ positive), and the anti-apodized case ($\gamma$ negative).

Furthermore, the rectangular geometry exhibits a very interesting property, resulting in significant attenuation along the diagonals $y = \pm x$. As early as in 2001, Nisenson & Papaliolios (2001) interpreted this property as a double apodization along both the $x$ and $y$ axes. Fig. 6, we present a zoom on the diagonal regions of the experimental results obtained with the interferometric apodization using a rectangular aperture. The left panel displays the case without apodization, the apodized case in the centre, and the anti-apodized case on the right.

To validate the apodization effect, we performed normalized cuts with the same maximum intensity along the diagonal regions on a logarithmic scale. These cuts are shown in Fig. 7, with the experimental images in blue and the theoretical predictions in red dashed lines. The experimental results show a contrast of approximately $10^{-3.5}$ for angular separation between $2\lambda/D$ and $3\lambda/D$, as well as a similar contrast for more distant separations of about $6\lambda/D$. Once again, a good agreement between theory and experiment can be observed.

The IAH technique applied to rectangular geometries offers a significant advantage over other apodization techniques, such as the one proposed by Carlotti et al. (2008). Our method allows for simultaneous apodization along both $x$ and $y$ axes, providing a two-dimensional (2D) solution without the necessity for further adjustments or modifications. In contrast, the method of Carlotti et al. (2008) achieves 2D circular apodization through the incorporation of complementary phase masks in both arms of the MZI. Yet, for rectangular apodization, their method is unidimensional, focusing exclusively on the $x$-axis. This implies that a secondary apodization stage would be requisite to encompass both $x$ and $y$ axes, which could introduce additional complexity to the system. On the other hand, our method requires only a single apodization stage, making it easier to implement and achieve similar or even superior performance compared to more complex apodization methods. Additionally, significant attenuation is observed in the diagonal regions, indicating that the IAH technique with a rectangular geometry offers a general and precise idea of the possibility of applying the IAH technique to any geometric shape of a telescope. For example, segmented apertures are similar to those used in large observatories like the TMT telescope that we used in the simulations shown in our previous article (Chafi et al. 2023).

## 3.2 Circular aperture

In the study conducted by El Azhari et al. (2005), a Michelson interferometer was used to experiment with one-dimensional apodization of a rectangular aperture. Subsequently, Azagrouze, El Azhari & Habib (2008) investigated two-dimensional apodization. Carlotti et al. (2008) performed an experimental validation of this apodization in the laboratory using a MZI, with converging and diverging lenses placed in the arms of the interferometer.

Following the mathematical formalism presented in our previous article (Chafi et al. 2023), we have extensively examined the theory associated with circular apertures, both unapodized and apodized, in the image plane. As a reminder, we introduced mathematical expressions for the transmission using a circularly symmetric top-hat function $\Pi\left(\frac{r}{R}\right)$, where $\Pi$ represents the circularly symmetric top-hat function with a radius $R$. This function value 1 inside the circle of radius $R$ and 0 outside. Furthermore, the amplitude associated with these apertures is given by $2\frac{J_1(\pi r)}{\pi r}$ (Born & Wolf 1980), and the apodization parameters $\gamma$ and $\eta$ are used to control the diameters and amplitudes of the beams, respectively. The transmittance is given by $\gamma\Pi\left(\frac{r}{\eta R}\right)$, and its associated amplitude is $2\gamma\frac{J_1(\pi\eta r)}{\eta\pi r}$. The analytical expression of the amplitude associated with the apodized pupil in the focal plane $P_1$ of the lens $L_s$ is expressed as follows:

$$\Psi(r, \gamma, \eta) = 2\frac{J_1(\pi r)}{\pi r} + 2\gamma\frac{J_1(\pi r \eta)}{\pi r \eta}, \qquad (4)$$

where $J_1$ is the first-order Bessel function.

The amplitude associated with the output in the opposite phase, displaying the anti-apodized case, in the final focal plane $P_1'$, can be expressed by equation (4) by replacing $\gamma$ with $-\gamma$.

The real parameters $\gamma$ and $\eta$ are used to control the diameters and amplitudes of the beams, respectively. This optimization facilitates the superposition of the two PSFs. As a result, the pedestals of the PSFs can be effectively compensated, leading to the achievement of apodization.

Before implementing the IAH technique, we conducted a sizing study to adapt the dimensions of the components (diffraction apertures and neutral density filters) to the available components in the market. This study was necessary to achieve the optimal performance of IAH in our experiment. As described in Article 1 (Chafi et al. 2023), we carried out an optimization of various parameters, including the diameters of the diffraction apertures and the densities of the neutral density filters. These optimizations were performed while considering the optimal values of the parameters $\gamma = 0.525$ and $\eta = 0.625$. Through this optimization, we achieved an energy concentration of 93.6 per cent in the central diffraction lobe, which represents an improvement compared to 84 per cent in the case without apodization. We also provided a sizing study to find standard component values that closely match the optimal values for users who do not have access to exact components. These standard values were obtained while considering practical constraints such as component availability in the market and manufacturing limitations.

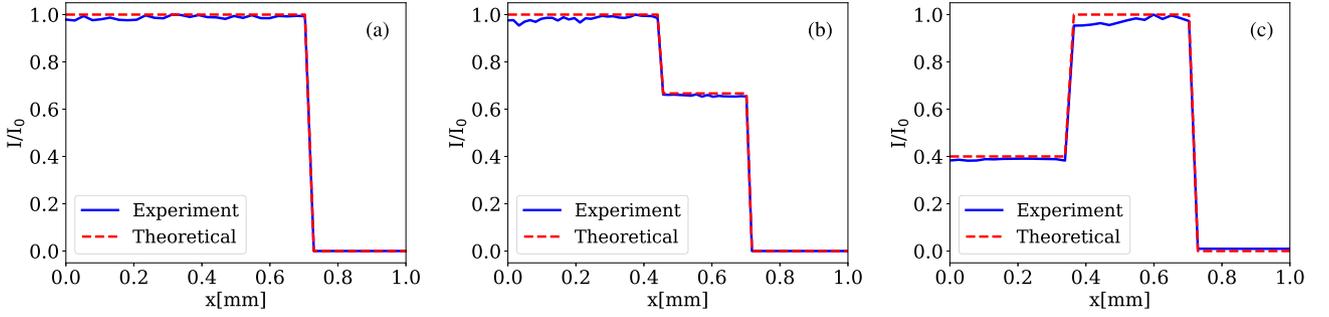

**Figure 4.** Transverse cross-sections along the horizontal axis, expressed in millimeters, of results in the pupil plane $P_2$ for a rectangular aperture, corresponding to Fig. 3(a), are presented in blue and compared to theoretical predictions represented by red dashed curves. The positions on the graph correspond to the non-apodized (a), apodized (b), and anti-apodized (c) pupils.

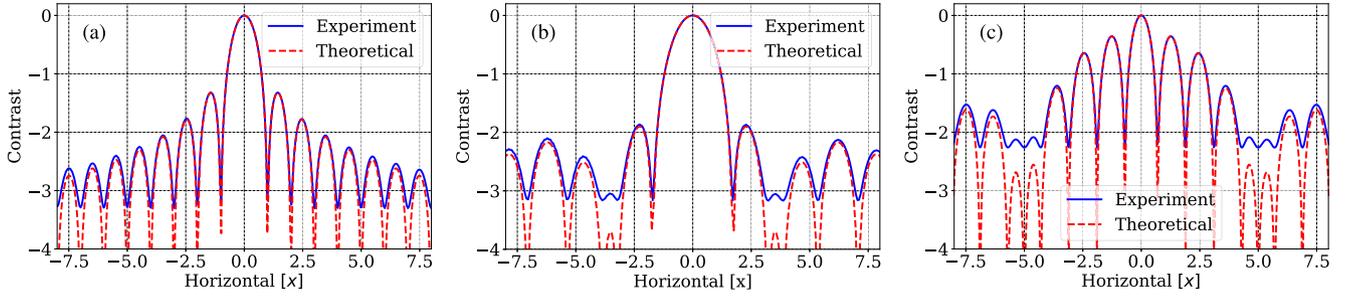

**Figure 5.** Theoretical predictions and corresponding experimental results are shown for the images presented in Fig. 3(b). The positions on the graph correspond to the non-apodized (a), apodized (b), and anti-apodized (c) PSF. The cross-sections along the $x$ axis, with the horizontal axis in units of $\lambda/D$, and the vertical axis representing the normalized intensity on a logarithmic scale.

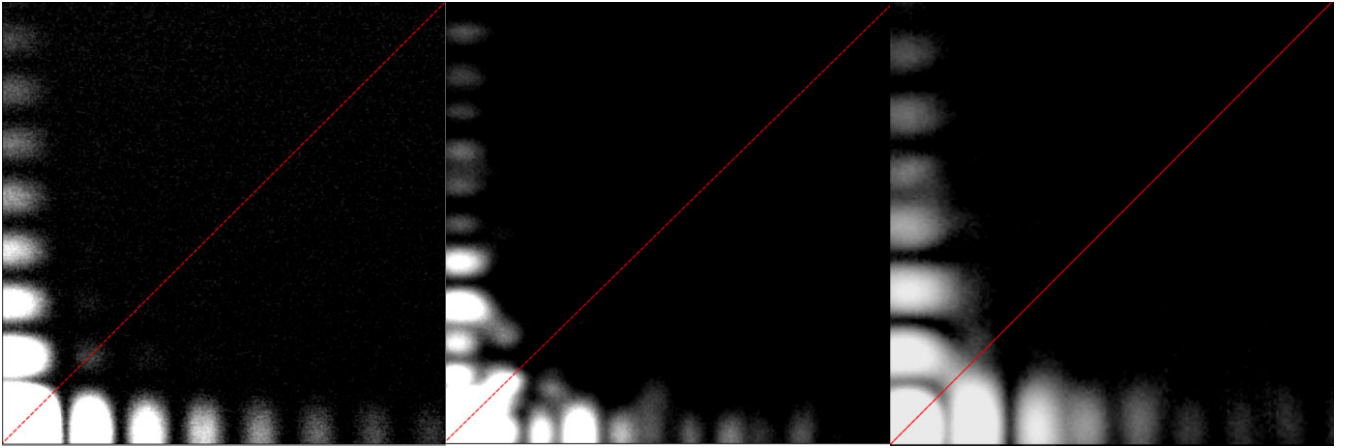

**Figure 6.** Zoom on the diagonal regions of the images in Fig. 3(b), with the same position, to display the effect of IAH technique in these zones.

To size our experiment, we adopted standard parameters ($\gamma = 0.498$ and $\eta = 0.6$) as presented in Section 3.3 of our previous article (Chafi et al. 2023). These values were chosen to achieve performance close to the optimal values. As for the diameter of the diffraction aperture $\mathcal{P}_1$, we opted for $\phi_1 = 1000\,\mu\text{m}$, a commonly available value in most optical component catalogs. Using this value and the homothety ratio ($\eta = \frac{\phi_2}{\phi_1}$), we obtained a diameter value for the second aperture $\mathcal{P}_2$ of approximately $600\,\mu\text{m}$. For the selection of the neutral density filters, we used specific densities to achieve a difference in density $\Delta D$ close to the optimal value of 0.15. Exactly, we chose neutral density filters with densities of $D_1 = 0.20$ and $D_2 = 0.04$, giving a $\Delta D$ of 0.16.

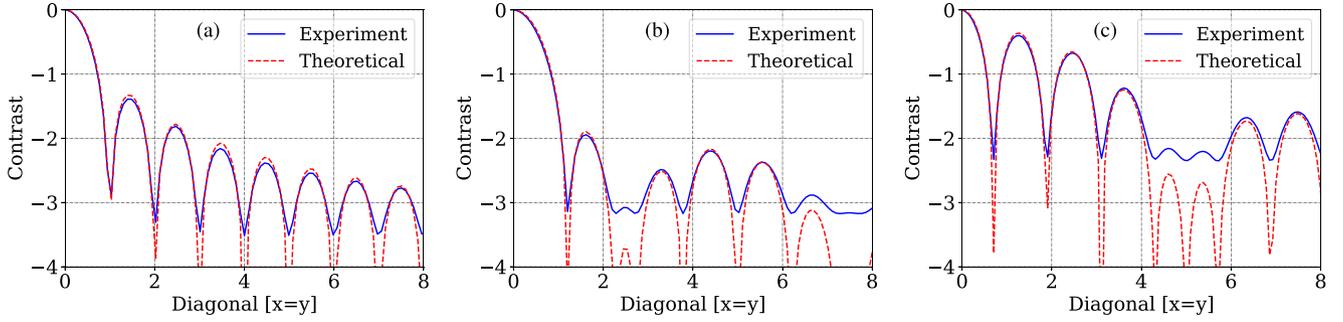

**Figure 7.** Diagonal cross-sections along $x = y$ from the results in the pupil plane P₁ for a rectangular aperture, corresponding to Fig. 6, are displayed in blue and compared to theoretical predictions on a logarithmic scale, represented by red dashed curves. The positions on the graph correspond to the non-apodized (a), apodized (b), and anti-apodized (c) PSF. The horizontal axis is scaled in units of $\lambda/D$, while the vertical axis represents the normalized intensity (or contrast) on a logarithmic scale.

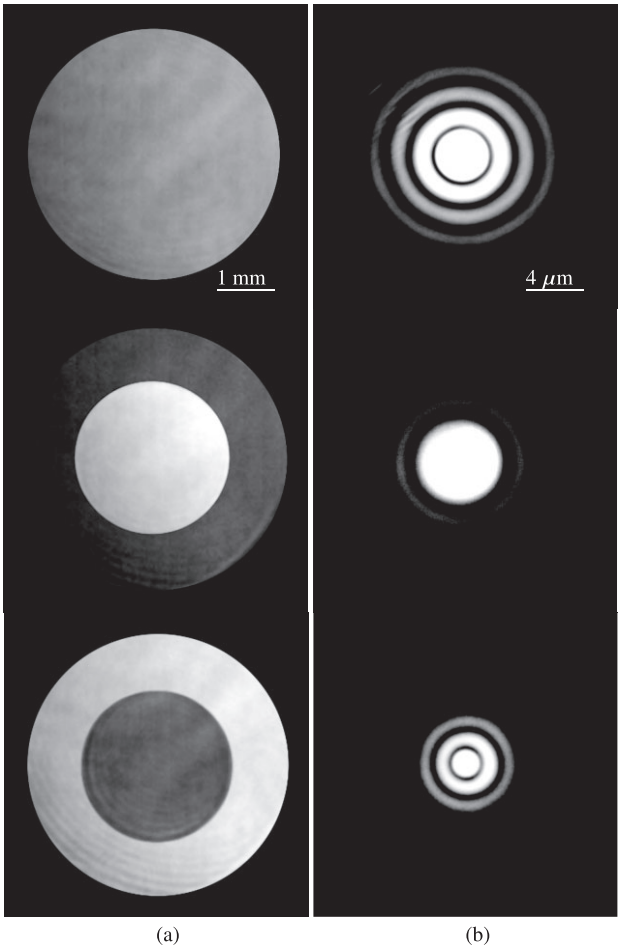

**Figure 8.** Apodization with circular symmetry. Column (a) depicts the pupil plane on a linear scale, while column (b) displays the focal plane on a logarithmic scale, for the cases of non-apodized (top), apodized (middle), and anti-apodized (bottom).

Fig. 8 presents the experimental results at the pupil plane [column (a)] and the focal plane [column (b)] considering three different situations: non-apodized (top), apodized (middle), and anti-apodized (bottom). These illustrations serve to qualitatively assess the impact of IAH on the phase outputs (apodized outputs) and the anti-phase outputs (or anti-apodized outputs) of the MZI. To obtain a non-apodized image, the light beam propagating along one of the arms of the MZI is blocked using a movable screen.

The results show good agreement with the simulation results. In comparison with the Airy disc, the apodized PSF exhibits an enlarged central core and a substantial reduction in secondary rings. Conversely, the anti-apodized PSF displays a slightly narrower central core, but with higher diffraction wings, particularly in the first two rings.

The radial profiles of the images and PSFs from Fig. 8 are shown in Figs 9 and 10. The experimental data are depicted in blue, while the theoretical projections are shown in red. In all cases, whether it is the non-apodized situation (a), apodized (b), or anti-apodized (c), can see a significant correspondence between the experimental data and the theoretical projections. Linear scales are used for the pupil images and logarithmic scales for the PSFs.

To improve the representation, all experimental curves are obtained from azimuthal averaging. This method highlights the overall trends of intensity profiles, making it particularly suitable for systems with radial symmetry. The principle of azimuthal averaging involves calculating the average of intensities along concentric circles centred on the interest point, providing a smoother and more robust representation of intensity variations.

The experimental results demonstrate a good agreement between theory and practice up to angular distances slightly greater than $7.5\lambda/D$. In addition to the expected enlargement of the central lobe of the PSF, a significant attenuation of the secondary rings is also observed.

In our comparative study on the IAH technique, we juxtaposed our experimental results with those derived from other apodization techniques. Our findings demonstrated an average contrast of approximately $10^{-2.5}$ at angular separations ranging from $2\lambda/D$ to $3\lambda/D$, and $10^{-4}$ between $6.5\lambda/D$ and $8\lambda/D$. In comparison, other apodization studies have reported an attenuation of around $10^{-2.5}$ at approximately $3\lambda/D$ (Carlotti et al. 2008). Our results align with the objectives of similar techniques that aim to attenuate the diffracted stellar intensity, as illustrated by Por (2017), Asmolova et al. (2018), and Zhang et al. (2018). These approaches, including ours, also target the optimization of performance for interferometric devices used in large telescopes. These comparisons underscore that our

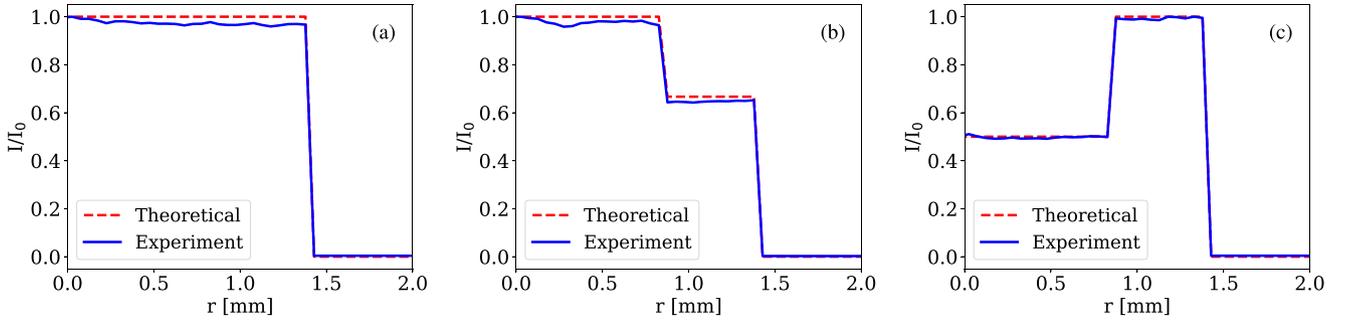

**Figure 9.** Comparison of azimuthal averages of the pupil (blue curves) from Fig. 8(a), and the theoretical predictions (dashed red curves). The positions on the graph correspond to the non-apodized (a), apodized (b), and anti-apodized (c) pupils. The horizontal axis is measured in millimeters, while the vertical axis represents the normalized intensity on a linear scale.

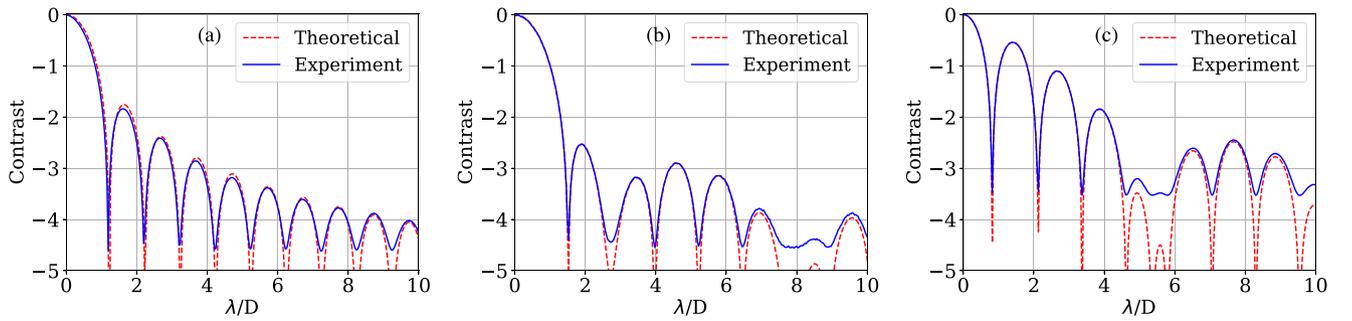

**Figure 10.** Alignment between theoretical predictions (dashed red) and experimental data (blue) for circular symmetry apodization. The positions of the curves correspond to the images from Fig. 8(b). The positions on the graph correspond to the non-apodized (a), apodized (b), and anti-apodized (c) PSF. The horizontal axis is measured in units of $\lambda/D$, while the vertical axis represents the normalized intensity on a logarithmic scale.

IAH approach offers commendable contrast performance, even at minor angular separations

### 3.3 Effects of neutral density filters

We have modified the transmission values $T$ of the neutral density filters to study their effect on our IAH technique. The transmission coefficients $T_i$ ($i = 1, 2$) of the neutral density filter are expressed as $T_i = 10^{-Di}$, where $D$ is the transmission coefficient of the neutral density filter given by $\Delta D = D_1 - D_2 = 2 \log_{10}(\frac{\gamma}{\eta})$.

Fig. 11 illustrates the evolution of the average azimuthal intensity in the final focal plane with respect to the transmission $T$ of the neutral density filters: the blue dashed curve at 97.2 per cent ($\gamma = 0.58$) and the green curve at 44.45 per cent ($\gamma = 0.4$) exhibit minimal attenuation and fail to achieve maximum contrast for angular separations. In contrast, the red curve at $T = 69.3$ per cent ($\gamma = 0.498$), which corresponds to the optimal value of $T$ with parameters $\gamma = 0.598$ and $\eta = 0.6$, shows a higher attenuation compared to the blue and green dashed curves while maintaining a high contrast in angular separations. In essence, the optimal value of $T$ strikes a favorable balance between filter transmission and image contrast. In this experiment, the value of $T = 69.3$ per cent offers a compelling compromise between attenuating diffracted light and preserving image contrast in angular separations. Therefore, this value of $T$ is considered the most optimal for this particular experiment.

### 3.4 Chromaticism

The effect of chromatic dispersion is a significant limitation in many optical applications, particularly in imaging and astronomical observations. To address this issue, various solutions have been proposed. One such solution was presented by Wynne (1979), who studied chromatism by using an achromatic system to produce an image with a focal ratio proportional to $1/\lambda$. Additionally, Aime (2005) proposed using prolate functions to make the Apodized Pupil Lyot Coronagraph (APLC) achromatic. The MZI has also been employed to achieve a more precise approximation of chromatic apodization, rather than relying solely on chromatic lenses or multiple glasses to compensate for the chromatic effect, as studied by Carlotti et al. (2008). Another solution was proposed by Aime et al. (2010), who introduced a new instrumental concept called the Achromatic Rotation-shearing Coronagraph (ARC) to eliminate chromaticity in astronomical observations. Similarly, Guerri et al. (2011) experimentally validated an achromatic apodizer in three bands: $H$ (1.6 μm), $J$ (1.191 μm), and $Y$ (1.063 μm).

While achieving achromatic apodization for all wavelengths remains a challenge, we evaluated the chromatic effect on the IAH technique by using three helium–neon laser sources at different wavelengths ($\lambda_0 = 543.5$ nm, $\lambda_1 = 594$ nm, and $\lambda_2 = 632.8$ nm). The average azimuthal intensity curves in the focal plane exhibit a shift in angular separation, as shown in Fig. 12, but with similar contrast levels for the three wavelengths tested, indicating that the chromatic effect on the central lobe is very weak. In other words, the performance of the IAH technique is insensitive to wavelength variations. This result is consistent with the nature of IAH, which is based on diffraction principles that affect all wavelengths in proportion to the aperture used. Furthermore, IAH's ability to provide similar performance over a wide range of wavelengths makes it highly advantageous for applications requiring imaging in different spectral bands.

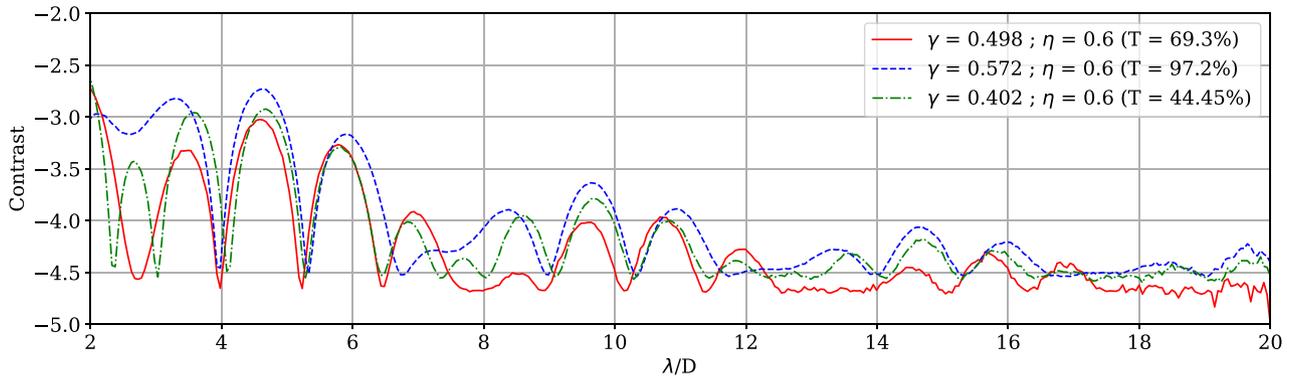

**Figure 11.** The azimuthal average in the final focal plane with different values of transmission $T$ is presented in logarithmic scale: dotted blue corresponds to $T = 97.3$ per cent, green represents $T = 44.45$ per cent, and red indicates the optimal value of $T = 69.3$ per cent.

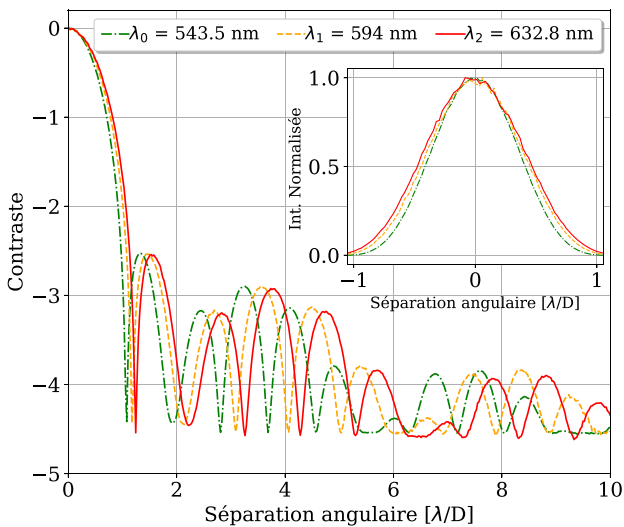

**Figure 12.** The average azimuthal intensities in the focal plane are displayed in logarithmic scale with different colours as follows: green for $\lambda_0 = 543.5$ nm, orange for $\lambda_1 = 594$ nm, and red dotted line for $\lambda_2 = 632.8$ nm. The effect of chromaticity on the central lobe is illustrated by the normalized intensity in linear scale displayed at the top.

Although IAH is relatively insensitive to wavelength aberrations, broad-band PSFs pose distinct challenges due to their rich spectral diversity. Indeed, when subjecting broad-band light to IAH, the risk of chromatic aberrations arises, potentially affecting the quality of the apodized PSF. However, our preliminary data, backed by the fundamental theory of IAH, confirm that the use of achromatic components, such as simple apertures and neutral density filters, allows IAH to reduce sensitivity to this spectral diversity. With this in mind, we plan to further validate our approach by experimenting with IAH using white light in our future studies. Finally, achromatism is a crucial property for astronomical observation instruments as it enables high-quality imaging across a broad range of wavelengths without the need for constant realignment of optics to compensate for wavelength variations.

## 4 CONCLUSIONS

The study presented in this article highlights the potential of the IAH technique for improving the performance of a coronagraph in terms of contrast at very small angular separations. Our experimental results are in good agreement with theoretical predictions, and the IAH method has proven to be competitive with other apodization techniques. We have also observed a significant attenuation of the secondary rings and a noticeable broadening of the central PSF lobe. The contrast achieved with the IAH technique is highly competitive, reaching up to $10^{-2.5}$ for angular separations between $2\lambda/D$ and $3\lambda/D$, and up to $10^{-4}$ for distances between $6.5\lambda/D$ and $8\lambda/D$.

Furthermore, our IAH approach with a rectangular geometry demonstrates a dual apodization effect in the diagonal regions as well as the apodization effect applied along both the $x$ and $y$ axes. Other techniques, as studied by Carlotti et al. (2008), only apply to a single axis, which complicates the application of apodization to both the $x$ and $y$ axes. In contrast, the IAH technique with a rectangular geometry offers a general idea of the possibility of applying this technique to telescopes with more complex geometric shapes, such as segmented telescopes used in large ground-based or space telescopes.

The IAH approach can be applied to any geometric shape of the telescope and has proven to be very useful in improving the performance of commonly used coronagraphs (e.g. APLC, Vortex, FQPM, etc.) in terms of average contrast at small separations. Although achieving achromatic apodization for all wavelengths remains a challenge in the direct imaging of exoplanets, the chromatic effect on the IAH technique has been studied using three Ne–He laser sources of different wavelengths. The results have shown that it is possible to achieve a similar but wavelength-shifted contrast due to the nature of diffraction. This indicates that the IAH technique could provide achromatic performance. We plan to further improve the technique by using polychromatic light in the next stage. The IAH technique shows great promise for high-contrast imaging and is comparable to other similar techniques, such as those using Michelson or MZI.

Moreover, it is interesting to note that the exploration of alternative aperture shapes, such as hexagonal or segmented, could be considered for the experimental implementation of the IAH technique. However, a thorough optimization study would be necessary to assess the advantages and challenges associated with these new aperture shapes in order to determine the optimal parameters ($\gamma$; $\eta$). Additionally, a procedure for acquiring new equipment would also be required to implement these alternative experimental configurations. Thus, the experimental realization of IAH with hexagonal or segmented apertures could be considered an objective of future work aimed at deepening the understanding of this technique and expanding its potential applications. In fact, in the experimental part of this study, we did not address either central obstruction apertures or segmented apertures. However, in previous work (Chafi et al. 2023), based on simulation calculations for IAH-apodized pupil responses, we demonstrated the significant potential of the IAH technique in reducing diffractive effects, both from the outer contour and the gaps between segment spaces. In this previous study, we used the values of optimized parameters determined for the case of a rectangular aperture in our simulations. It might be worthwhile to determine optimized parameters specific to other geometries. The principle remains the same, but the calculation procedures would need to be modified. Transitioning to experimentation will require designing the experimental set-up before proceeding with specific adjustments and measurements.

The IAH technique holds promising perspectives for celestial observations or tests, especially when combined with a coronagraph, allowing significant contrast enhancement and improved performance. However, it is essential to acknowledge that realizing such observations entails several complex challenges that need to be addressed. A careful examination of the technical and logistical requirements is crucial to implement this experimental manipulation within an observational framework, especially when using a telescope with a minimum diameter of 6 metres to gather sufficient starlight and achieve the resolution needed for exoplanet detection. This involves considerations such as the availability of appropriate equipment, precise planning of experimental conditions, and managing potential technical limitations. It should be noted that all these aspects will be the subject of future studies.

In summary, the IAH technique presented in this article is a promising approach to enhance the contrast of a coronagraph and can be applied to any geometric shape of a telescope. In the next stage, we plan to further improve the IAH technique by using multiple cascaded stages to increase the contrast at very short separations.

## DATA AVAILABILITY

The data underlying this article will be shared on reasonable request to the corresponding author.

This paper has been typeset from a TeX/LaTeX file prepared by the author.